\documentclass[sigconf]{acmart}

\usepackage{xcolor}
\usepackage{algorithm}
\usepackage{algorithmic}
\usepackage{natbib}

\AtBeginDocument{%
  \providecommand\BibTeX{{%
    \normalfont B\kern-0.5em{\scshape i\kern-0.25em b}\kern-0.8em\TeX}}}

\copyrightyear{2024}
\acmYear{2024}
\setcopyright{acmlicensed}\acmConference[AST '24]{5th ACM/IEEE International Conference on Automation of Software Test (AST 2024)}{April 15--16, 2024}{Lisbon, Portugal}
\acmBooktitle{5th ACM/IEEE International Conference on Automation of Software Test (AST 2024) (AST '24), April 15--16, 2024, Lisbon, Portugal}
\acmDOI{10.1145/3644032.3644458}
\acmISBN{979-8-4007-0588-5/24/04}

\begin{document}

\title{Testing for Fault Diversity in Reinforcement Learning}

\author{Quentin Mazouni}
\affiliation{%
  \institution{Simula Research Laboratory}
  \city{Oslo}
  \country{Norway}
}
\email{quentin@simula.no}

\author{Helge Spieker}
\affiliation{%
  \institution{Simula Research Laboratory}
  \city{Oslo}
  \country{Norway}
}
\email{helge@simula.no}

\author{Arnaud Gotlieb}
\affiliation{%
  \institution{Simula Research Laboratory}
  \city{Oslo}
  \country{Norway}
}
\email{arnaud@simula.no}

\author{Mathieu Acher}
\affiliation{%
  \institution{Univ Rennes, Inria, INSA Rennes, CNRS, IRISA}
  \city{Rennes}
  \country{France}
}
\email{mathieu.acher@irisa.fr}

\begin{abstract}
Reinforcement Learning is the premier technique to approach sequential decision problems, including complex tasks such as driving cars and landing spacecraft.
Among the software validation and verification practices, testing for functional fault detection is a convenient way to build trustworthiness in the learned decision model.
While recent works seek to maximise the number of detected faults, none consider fault characterisation during the search for more diversity.
We argue that policy testing should not find as many failures as possible (e.g., inputs that trigger similar car crashes) but rather aim at revealing as informative and diverse faults as possible in the model.
In this paper, we explore the use of quality diversity optimisation to solve the problem of fault diversity in policy testing.
Quality diversity (QD) optimisation is a type of evolutionary algorithm to solve hard combinatorial optimisation problems where high-quality diverse solutions are sought.
We define and address the underlying challenges of adapting QD optimisation to the test of action policies.
Furthermore, we compare classical QD optimisers to state-of-the-art frameworks dedicated to policy testing, both in terms of search efficiency and fault diversity.
We show that QD optimisation, while being conceptually simple and generally applicable, finds effectively more diverse faults in the decision model, and conclude that QD-based policy testing is a promising approach.
\end{abstract}



\maketitle

\section{Introduction}

In the last decade, Reinforcement Learning (RL) combined with neural networks (NNs) was shown to be able to effectively solve complex sequential decision problems in various fields, such as planning, game playing or system control~\cite{doi:10.1126/science.aar6404, DBLP:journals/corr/abs-1908-01362, DBLP:journals/corr/abs-2007-06702}.
Deployment of such action policies in real-world applications involves strong software validation and verification practices.
Among them, testing, which trades exhaustiveness for efficiency helps build trust and confidence in the decision model.
To that regard, a lot of work to test learnt policies have recently emerged~\cite{10.1145/3533767.3534388, 9712397, 10172658, zolfagharian2023searchbased}.
Some of them~\cite{10.1145/3533767.3534388, zolfagharian2023searchbased} further investigate the test results after the search.
For example, \citeauthor{10.1145/3533767.3534388}~\cite{10.1145/3533767.3534388} retrain the policy and~\citeauthor{zolfagharian2023searchbased}~\cite{zolfagharian2023searchbased} extract interpretable rules from a Decision Tree to characterise the fault-revealing inputs.
However, none of them proposes to consider how the policy under test solves (or fails) the problem at each test case and look for diversity.
Instead, they all maximise the number of faults found, whatever they reveal or mean.

Still, fault diversity is important: to improve the model, to assess the range of possible incorrect decisions (policy explainability) and one can also see the diversity as a test coverage measure, the latter being especially difficult to assess in NN-based policies.
That way, stakeholders could accurately assess the safety of the NN-based policy and eventually build trust and trustworthiness.
Fortunately, an evolutionary search technique known as Quality Diversity (QD, or Illumination) addresses this very same issue.

Indeed, QD optimisation finds both diverse and high-quality solutions for a given task.
To do so, diversity is not computed in the search space but rather in a behaviour space that describes how a solution actually solves the task.
Typical QD applications are in robotics, where the objective is to find the best action policies (i.e., that successfully solve the task, like getting out of a maze) while discovering as many \textit{behaviourally} different policies as possible, e.g., how the robot leaves the maze.

In this paper, we propose to address the problem of fault diversity in policy testing with Quality Diversity.
In other words, we investigate if QD optimisation can find diverse faults in trained policies.
Especially, we define and address the underlying challenges of adapting QD optimisation to policy testing.
We compare how two existing QD optimisers solve the subsequent policy testing task as a QD problem  to a state-of-the-art framework dedicated to policy testing, both in terms of search efficiency and fault diversity.
Our results show that QD-based testing finds diverse solutions, without additional test budget cost.
The contributions of this paper are thus:
\begin{itemize}
    \item We propose the first reformulation of policy testing as a Quality Diversity optimisation problem.
    \item We implement our method with two classical Illumination algorithms.
    \item We compare the two QD-based frameworks to a SOTA framework dedicated to policy testing on three use-cases.
\end{itemize}
The rest of the paper is organised as follows.
Section~\ref{sec:rel_work} describes the current policy testing techniques and positions our study.
Section~\ref{sec:background} introduces Reinforcement Learning for decision making and Quality Diversity optimisation.
Section~\ref{sec:method} presents the challenges to solve the problem of fault diversity in policy testing in the Quality Diversity framework.
Section~\ref{sec:empirical_evaluation} describes the empirical evaluation of our QD-based policy testing framework implemented with 2 QD optimisers to a dedicated policy testing technique.
We discuss the current limitation in Section~\ref{sec:threats}.
Eventually, Section~\ref{sec:conclusion} concludes the paper and draws some perspectives.

\section{Related Work}
\label{sec:rel_work}
Policy testing has been recently addressed in numerous ways~\cite{mazouni2023}.
In the following, we describe the different testing objectives studied alongside the corresponding testing techniques proposed.

\citeauthor{zolfagharian2023searchbased}~\cite{zolfagharian2023searchbased} test RL policies with a genetic algorithm~\cite{10.2307/24939139}, where test cases are individuals of a population.
Here, the individuals are episodes and their genes, state-action pairs.
Starting from historical data (e.g., training data of the policy tested), the search consists of evolving the population to generate faulty episodes which are likely to be consistent with the policy under test (i.e., the trajectories match policy's decision).
While this approach lets it avoid executing the policy, it also requires that resulting faulty episodes be validated with respect to the policy.
~\citeauthor{9712397}~\cite{9712397} and \citeauthor{10172658}~\cite{10172658} investigate active policy testing, which consists of dynamically changing the simulator (the policy under test interacts with) during executions.
As such, the test cases are defined as the sequences of the environmental changes applied to the simulator.
They both turn the search problem into a RL task, where an agent learns to perturb the simulation to generate hazard decisions in the policy.
More precisely, \citeauthor{9712397}~\cite{9712397} address the case of an autonomous driving system (where the possible modelled perturbations include changing weather conditions and dynamics of pedestrians and vehicles -- e.g. their position or velocity --), while \citeauthor{10172658}~\cite{10172658} consider the case of multiple testing requirements (i.e., many-objective search).
\citeauthor{https://doi.org/10.48550/arxiv.2205.04887}~\cite{https://doi.org/10.48550/arxiv.2205.04887} look for states that trigger unsafe decisions, called boundary states.
The crucial difference with all the other methodologies is that they do not look for those boundary states but, rather, retrieve the latter from the state space explored by an initial backtracking-based, depth-first search for a solution of the decision-making problem.
This approach can thus be computationally expansive (depending on the difficulty of the decision task) and provides no guarantee of finding boundary states from the search.

Fuzzing frameworks have also been recently proposed~\cite{10.1145/3533767.3534388, Steinmetz_Fišer_Eniser_Ferber_Gros_Heim_Höller_Schuler_Wüstholz_Christakis_Hoffmann_2022, 10.1145/3533767.3534392}.
\citeauthor{10.1145/3533767.3534388}~\cite{10.1145/3533767.3534388} consider seeds as initial situations of the decision-making task.
Similarly to genetic searches, the search space is explored by mutating used seeds.
Even though the search does not look for diversity, it accounts for novelty by maintaining the pool of seeds which produce uncovered state sequences.
Precisely, they compute the likelihood of the latter (collected after each execution) with Gaussian Mixture Models~\cite{10.2307/2982840} and keep a seed only if the likelihood is lower than a defined threshold.
As for~\cite{Steinmetz_Fišer_Eniser_Ferber_Gros_Heim_Höller_Schuler_Wüstholz_Christakis_Hoffmann_2022} and \cite{10.1145/3533767.3534392}, they investigate the \textit{bug confirmation} problem for NN-based action policies, which corresponds to finding \textit{avoidable} failures.
To bypass the oracle problem in such a situation, \citeauthor{Steinmetz_Fišer_Eniser_Ferber_Gros_Heim_Höller_Schuler_Wüstholz_Christakis_Hoffmann_2022}~\cite{Steinmetz_Fišer_Eniser_Ferber_Gros_Heim_Höller_Schuler_Wüstholz_Christakis_Hoffmann_2022} use heuristics known in classical AI planning, and \citeauthor{10.1145/3533767.3534392}~\cite{10.1145/3533767.3534392} rely on metamorphic testing.
To do so, the authors design the metamorphic operations around state relaxation, a well-studied concept also taken from the AI planning community.
Their idea is that a relaxed version of a given environment should represent an easier problem than the original one.
Therefore, the policy under test contains a bug if it solves the original problem but fails to solve its ``relaxed'' counterpart.
Besides, \citeauthor{10.1145/3180155.3180220}~\cite{10.1145/3180155.3180220} and \citeauthor{DBLP:journals/corr/PeiCYJ17}~\cite{DBLP:journals/corr/PeiCYJ17} consider white-box testing of image input-based NNs.
In their work, the objective is to find behaviour inconsistencies, which is approached as a neuron-coverage-guided greedy search.
They differ in their test oracles: \citeauthor{10.1145/3180155.3180220}~\cite{10.1145/3180155.3180220} use metamorphic testing~\cite{Chen1998} to check that the model tested outputs the same action for morphed images, while \citeauthor{DBLP:journals/corr/PeiCYJ17}~\cite{DBLP:journals/corr/PeiCYJ17} rely on differential testing~\cite{McKeeman1998DifferentialTF} (i.e., several NNs are simultaneously tested and inconsistencies are detected when their decisions differ).

\section{Background}
\label{sec:background}

In this section, we introduce Reinforcement Learning to approach sequential decision-making and Quality Diversity optimisation.

\subsection{Reinforcement Learning for sequential decision-making}

Informally, sequential decision-making refers to tasks that can be solved by any decision model in a step-by-step manner and which accounts for the dynamics of the environment~\cite{FrankishRamsey2014}.
This definition is very broad, and in the following we consider tasks that involve a single decision entity.
As such, solving a sequential decision-making problem consists of initialising the world (or environment) to a particular starting situation and letting the decision model (or agent) interact with the former (e.g., simulations) through a step-wise observation-decision-action process until a final state is reached: if the latter is satisfying, the agent has solved the task; otherwise, the agent fails.
A typical example is the case of path planning in robotics, where the agent is expected to safely reach a given position from an initial situation.
Formally, those kind of problems are formulated as Markov Decision Processes (MDPs), defined as 4-tuples $\langle S,A,R,T\rangle$ where:
\begin{itemize}
    \setlength{\parskip}{2pt}
    \item $S$ is a set of states. Referred to as the observation space, it corresponds to what the agent can know about its environment.
    \item $A$ is the set of actions. Referred to as the action space, it specifies how the agent acts on its environment.
    \item $R:S\times A\mapsto\mathbb{R}$ is the reward function. It reflects the agent’s performance by associating any pair of state-action with a numerical value.
    \item $T:S\times A\times S\mapsto[0,1]$ is the transition function, which is a probability distribution over the observation and the action space. It depicts which state the environment will transit to after an action is executed. The function is not known by the agent and governs the environment's dynamic.
\end{itemize}
Solutions to MDPs are called policies and noted $\pi$, which are functions mapping from the observation space $S$ to the action space $A$.
A common approach to training policies is Reinforcement Learning, a sub-field of Machine Learning which consists in learning from rewards/penalties~\cite{sutton2018reinforcement}.
Precisely, RL learns an optimal, or nearly-optimal, policy $\pi:S\times A\mapsto[0,1]$ that maximises the total expected discounted cumulative reward $R_t=\sum_{t>0} \gamma^{t-1}r_t$, where $0<\gamma\leq1$ is the discount factor.
This parameter controls how the agent takes into account the future rewards.
Precisely, a small value encourages the agent to maximise short-term rewards, whereas high values (usually close to 1) lead the agent to focus on maximising long-term rewards.

In this work, we consider black-box testing, i.e., without access to the internals of the policy nor the simulator, of \textit{deterministic} decision models, changing thus the previous definition to $\pi:S\mapsto A$.
In the following, we introduce Quality Diversity optimisation.

\subsection{Quality Diversity}
\label{sec:qd}

Informally, Quality Diversity optimisation stems from the evolutionary algorithms, but provides a shift in methodology by not only considering the maximisation of a fitness function, but explicitly targeting the discovery of diverse solutions as characterised by their \textit{behaviour}, i.e. the way a problem is solved.
Formally, it assumes the objective function $f$ now returns a fitness value $f_x$ and a behavioural descriptor $b_x$ for any parameter $x$, i.e., $f_x, b_x \leftarrow f(x)$.
As previously mentioned, the behavioural descriptor describes how the solution solves the problem, while the fitness value quantifies how well it is solved.
If we assume that we want to maximise the fitness function and define the behaviour space as $B$, then the goal of QD optimisation is to find for each point $b \in B$ the parameters $x \in X$ with the maximum fitness value:
\[\forall b \in B, x^* =\arg\max_{x}f_x \mid b_x = b\]
The goal of QD is thus to return a collection of behaviourally distinct solutions, whose individuals are the best performers (also called \textit{elites}) in their local behaviour area (or behaviour \textit{niches}).

The first method in the QD paradigm is \textit{Novelty Search} (NS)~\cite{Lehman2008ExploitingOT,lehman2011abandoning}, which entirely abandons the search for the objective function and only explores behavioural novelty.
While later developments in QD reconsider the inclusion of the objective function into the search by competition of behaviourally similar solutions with different quality~\cite{10.1145/2739480.2754736,10.1145/2001576.2001606}, we see a parallel between the pure search for novelty and fault-triggering inputs for policies.
In the search for fault-triggering inputs, there is no clear objective function to maximise besides the binary test verdict of a successful or failing episode. There is no indication of the closeness to a failure, even though some existing works introduce surrogate objectives to help guide the search~\cite{zolfagharian2023searchbased}.
We therefore consider novelty search as a potentially interesting approach to explore the behaviour space without the necessity for guidance through an objective function.

In its recent variants, QD tends to discover as many diverse behaviours as possible, while improving their elite, i.e., the individual with the highest quality in that niche.
The result of QD optimisation is a set of solutions, which is usually referred to as \textit{collection}, \textit{archive} or \textit{container} and is structured through the niches of the behaviour space.
During the optimisation process, the collection is filled with any candidate whose behaviour is novel enough and lets the latter compete with the current collection's elite if its behaviour is deemed to belong in a niche already covered.
As such, the collection of any QD algorithm structures its search, since it defines the behavioural neighbourhood of the parameters evaluated (i.e., how to consider if two solutions have close enough behaviours to belong to the same behavioural area or niche).

In the QD literature, one can distinguish two types of collections, namely structured and unstructured ones.
In the former case, the behaviour space is discretised with a segmentation pattern into a grid, and each cell represents a niche (or behaviour descriptor location).
This approach is for example implemented in the MAP-Elites algorithm~\cite{mouret2015illuminating}, one of the most used QD optimisers.
Here, the collection is a regular grid and the search aims at filling every cell of that grid with the best possible solution.
On the other hand, unstructured archives do not define niche locations before the optimisation starts and relies on distance thresholds and/or local densities to assess the behavioural similarities and neighbourhoods between solutions.

\citeauthor{7959075}~\cite{7959075} propose a unified definition of QD optimisation that we follow to introduce the general, high-level optimisation process in QD (see Algorithm~\ref{alg:qd}).
In their formulation, algorithms vary depending on (i) the type of container (i.e., how the data is gathered and ordered into a collection);
(ii) the selection operator (i.e., how the solutions to be modified in the next generation are selected)
and (iii) the type of scores computed for the container and the selection operator to work.
Given those parameters and operators, the execution of a QD algorithm follows a 4-step loop until the budget is consumed:
\begin{itemize}
    \item A new batch of candidates is produced from the individuals selected by the selection operator.
    \item The candidates are evaluated and their performance and descriptors, recorded.
    \item Each candidate is possibly added to the container (according to the solutions already in the collection).
    \item The scores maintained by the container are eventually updated (if needed). Common scores are the novelty, the local competition, or the curiosity score.
\end{itemize}
For more information about the different variants studied and the most widely used scores, see~\cite{7959075}.

\definecolor{change}{RGB}{220, 50, 100}
\definecolor{original}{RGB}{100, 140, 220}
\definecolor{gray}{RGB}{150, 150, 150}

\begin{algorithm}
\renewcommand{\algorithmicrequire}{\textbf{Input:}}
\renewcommand{\algorithmicensure}{\textbf{Output:}}
\renewcommand{\algorithmiccomment}[1]{\textit{\textcolor{gray}{#1}}}
\newcommand{\LONGCOMMENT}[1]{\textit{\textcolor{gray}{$\triangleright$ \: #1 \hfill $\triangleleft$}}}
\caption{4-step QD-algorithm for Policy Testing}
\label{alg:qd}
\begin{algorithmic}[1]
\REQUIRE $N$: iteration budget, $N_{init}$: number of initial iterations
\ENSURE $A$: archive of diverse and high-performing solutions
\STATE $A \gets \emptyset$ \COMMENT{\hfill $\triangleright$ empty archive of solutions}
\FOR{$I \gets 0$ to $N$}
    \STATE \LONGCOMMENT{start with random parents and offspring}
    \IF{$I < N_{init}$}
        \STATE $X_{parents} \gets \text{random\_solutions()}$
        \STATE $X_{offspring} \gets \text{random\_solutions()}$
    \ELSE
    \STATE \LONGCOMMENT{1. select individuals from the archive and/or the previous batch}
        \STATE $X_{parents} \gets \text{select}(A, X_{offspring})$
        \STATE \LONGCOMMENT{2. create randomly modified copies of $X_{parents}$ (mutation and/or crossover)}
        \STATE $X_{offspring} \gets \text{variation}(X_{parents})$
    \ENDIF
    \FOR{$x \in X_{offspring}$}
    \STATE \LONGCOMMENT{3. record the behaviour and quality of the candidate}
        \STATE \textcolor{original}{$b_x, f_x  \gets f(x)$}
        \STATE \textcolor{original}{$x.\text{store\_score}(b_x, f_x)$}
      \STATE \LONGCOMMENT{\small{3*. initialise $MDP$ with $x$ and characterise $x$ with (1) the behaviour and fitness of $\pi$; (2) the test oracle result}}
        \STATE \textcolor{change}{$b_x, f_x, o_x \gets Evaluate(MDP, \pi, x)$}\label{change1}
        \STATE \textcolor{change}{$x.store\_score(b_x, f_x, o_x)$}\label{change2}
        \STATE \LONGCOMMENT{4. attempt to add the candidate to the archive (local competition)}
        \STATE $\text{attempt\_to\_add}(x, A)$
        \STATE $\text{update\_scores}(\text{parent}(x), A)$ \LONGCOMMENT{\small{parent's scores might be updated}}
    \ENDFOR
    \STATE \LONGCOMMENT{\small{possibly update the scores of all solutions (e.g., curiosity scores)}}
    \STATE $\text{update\_scores}(A)$
\ENDFOR
\RETURN $A$
\end{algorithmic}
\end{algorithm}

\section{QD Optimisation for Policy Testing}
\label{sec:method}

This section describes the challenges in optimising policy testing for fault diversity with Quality Diversity and how we address them.
We consider a black-box setting, where neither the $MDP$ nor the policy under test $\pi$ can be directly inspected, but only inputs and outputs can be observed.
The only assumption we take is, that it is possible to instrument the $MDP$'s simulator with an initial state and random seed.

\paragraph{Solution Behaviour}

Since we aim to find fault-triggering initial states of the MDP (under which $\pi$ fails), the search space corresponds to the parameter (or input) space of the MDP's simulator.
The exact definition of such parameter spaces depends on the implementation of the simulator and the decision-making problem (see Section~\ref{sec:empirical_evaluation}).
Therefore, at first glance, the search space does not come with any behavioural definition directly: parameters like gravity or objects' positions do not \textit{behave} in a specific way.
This is not an issue though, as we can still use the traditional behaviour space definition of QD.
Indeed, we similarly want to characterise how the policy under test solves the MDP.
The main difference is that in QD, the solutions evaluated are policy's parameters (since the goal is to find good policies) whereas in our reformulation, $\pi$ is the same and its behaviours depends on the initial scenario described by the solution evaluated.
This allows us to leverage the QD's literature richness of behavioural policy analysis and use already proposed behaviour space definitions (see Section~\ref{sec:empirical_evaluation}) to describe the solutions.
That way, finding diverse solutions will exercise as many different policy behaviours as possible and, similarly, the fault-triggering ones will imply diverse hazard decisions.

\paragraph{Solution Quality}\label{subsec:quality}

The second challenge in the adaption of QD for policy testing lies in the definition of the solutions' quality (or fitness).
Indeed, in policy testing the quality of a test case (i.e., the evaluation of an execution) boils down to the boolean value of the test oracle (i.e., $\pi$ solves the task or fails), which is not enough informative.
We instead define the quality of solutions as the accumulated reward of the policy under test (like in common QD applications, i.e., as if we were finding policies) and define the optimisation task of QD with minimisation (i.e., accumulated reward minimisation).

\paragraph{Assumptions}

In this first work, we fix the randomness effects during the simulation, letting thus the $MDP$, its simulator and all their executions be deterministic.
As such, every input test (i.e., solution) generates a single trajectory and thus, a single behaviour (and test result).

\paragraph{QD-based Policy Testing}

Thanks to the flexibility of Quality Diversity, our proposal to optimise a fault-triggering simulator's inputs search for policy testing with QD involves only few changes to the high-level framework proposed by \citeauthor{7959075}~\cite{7959075} as shown in Algorithm~\ref{alg:qd}.
The modifications are highlighted at lines~\ref{change1}-\ref{change2}, which mostly consists of changing how solutions are evaluated.
First, a solution $x$ is used to initialise the $MDP$'s simulator.
Then, we let $\pi$ solve/fail the decision task and characterise $s$ with (1) the behaviour and fitness of $\pi$; (2) the test oracle result.
After the search, fault-triggering solutions (faults for short) can be retrieved from the archive by filtering it with the test oracle results.

\section{Experimental Evaluation}
\label{sec:empirical_evaluation}

\subsection{Research Questions}
To evaluate QD optimisation for fault diversity in policy testing, we conduct experiments to answer three research questions (RQs):

\begin{enumerate}
    \item[\textbf{RQ1}] How efficient is QD optimisation compared to dedicated policy testing techniques?
    \item[\textbf{RQ2}] How does QD optimisation improve diversity?
    \item[\textbf{RQ3}] How does the behaviour space definition impact the performance of QD-based policy testing?
\end{enumerate}

Answering the first two research questions will let us assess the benefits and cost of prioritising diversity.
Here, we compare the number of faults revealed and their diversity among QD-based testing and testing without consideration of behaviours.
Finally, the last research question investigates the impact of the definition of the behaviour space, a key configuration parameter for QD optimisation.
We expect that the number of faults found by QD-based testing still be competitive and that it improves fault diversity by accounting for the behaviour of the policy under test.

\subsection{Experiments}\label{subsec:exp_setup}
To answer the RQs, we conduct experiments with three standard environments~\cite{towers_gymnasium_2023}: Lunar Lander, Bipedal Walker and Taxi.
We compare Random Testing with MDPFuzz~\cite{10.1145/3533767.3534388}, a recent black-box policy testing technique for MDPs and two implementations of our QD-based testing framework.
The first one uses the QD optimiser MAP-Elites~\cite{mouret2015illuminating}, and the second one, Novelty Search~\cite{Lehman2008ExploitingOT}.
Random Testing will assess the difficulty of the testing task and act as a baseline to compare the other methodologies.
We choose the policy-dedicated testing framework MDPFuzz since it addresses complex environments (which has not been shown by other frameworks such as STARLA~\cite{zolfagharian2023searchbased}) and drives its fuzzing search towards uncovered state sequences, accounting thus for diversity.
Finally, as part of our QD-based testing framework, we study MAP-Elites (ME) since it is one of the first and most conceptually simple illumination algorithms, while Novelty Search (NS) will let us assess the relevance of accounting for the quality of the executions, as this algorithm emphasises diversity only.

\subsubsection{Environments}\label{subsubsec:ll}

The three selected environments are commonly used benchmarks in the RL literature.

\paragraph{Lunar Lander}\label{ll}
This control problem consists in safely landing a spacecraft.
We chose this environment since it has been used in QD optimisation and RL.
In particular, a behaviour space has already been studied~\cite{pyribs_lunar_lander}.
The spacecraft always starts at the top centre of the space and, similarly, the landing pad is always at the centre of the ground.
The initial situations differ in the shape of the landscape (around the landing pad) and the initial force applied to the spacecraft.
The policy controls the main and orientation engines of the spacecraft.
Precisely, there are four possible actions: do nothing, fire the left orientation engine, fire the main engine and fire the right orientation engine.

\paragraph{Bipedal Walker}\label{bw}
This problem consists in piloting a 4-joint walker robot across an uneven landscape composed of obstacles like steps, pits, and stumps.
We chose this environment to follow the evaluation of MDPFuzz~\cite{10.1145/3533767.3534388} (enabling thus a fair comparison), but also because behaviour spaces have already been proposed~\cite{10.1145/3377929.3389921}.
The impact of these definitions on the results are studied in RQ3.
In this problem, the walker always starts at the same position.
The initial situations differ in the shape of the landscape (positions of the steps, the stumps and the pits).
The action space is continuous.
Precisely, the action of the policy is the motor speed values at the 4 joints of the robot, which are localised at the hip and its knees.

\paragraph{Taxi}\label{tt}
In this classical environment, the policy navigates in a grid world to pick up passengers and safely drop them off at their destinations~\cite{dietterich2000hierarchical}.
Every test initiates a particular initial situation as the position of the passenger, the position of the taxi and the passenger's destination.
The six actions possible by this policy are the next taxi's direction (going north, south, east or west) and interactions with the passenger (pick it up, drop it off).
We use a version of the Taxi environment with an enlarged 18x13 map, thus disabling the simple enumeration of all MDP's possible states for the standard 5x5 grid.

\subsection{Metrics}

To answer RQ1, we measure what we call the {\it test efficiency} as the number of distinct faults found over time.
To answer RQ2, we study the diversity of testing (i.e., how a test methodology exercises the policy under test) and diversity of the faults.
We consider two metrics to measure diversity.
First, we compare the behaviour coverage, that is: how many behaviours are discovered during testing.
To do so, we follow the QD literature and count the number of bins filled in each result archive.
This archive corresponds to the regular grid used by MAP-Elites.
For fault diversity, only the bins filled by at least one fault-triggering solution are counted.
To complement this method space-based measure, we also analyse the diversity with the final states of the simulations.
The idea of the final state comparison is to make the result analysis possibly more accessible and not only do a comparison in the method space (i.e., behaviour space).
Indeed, the definition of the behaviour space is domain-dependent and can therefore vary.
Complementing thus the behaviour space coverage with final state diversity analysis will provide us with conclusions that are not bound to how behaviours are actually computed from the trajectories.
We report final state diversity as the average distances of the 3 nearest neighbours in the solution sets, since this metric captures the sparseness of a data set (i.e., if the set is composed of dense points or if the points are smoothly distributed).
Similarly as the first metric, fault diversity only considers the final states of fault-revealing executions (failure states).
Eventually, to answer RQ3, we compare the effect of four behaviour spaces for the Bipedal Walker use-case on all the previous metrics mentioned above.

\subsection{Implementation}\label{subsec:impl}

We run all methods with a budget of 5000 tests, and an initialisation phase of 1000 for MAP-Elites and MDPFuzz (following the evaluation in~\cite{10.1145/3533767.3534388}).
For Novelty Search, we use a population size of 100 and let the search iterate for 50 iterations.
The result archives used to collect data rely on a regular grid of 50x50 bins.
All the experiments were executed on a Linux machine (Ubuntu 22.04.3 LTS) equipped with an AMD Ryzen 9 3950X 16-Core processor and 32GB of RAM.
We accounted for randomness effects by repeating all the experiments with 10 seeds and report the median results as well as the first and third quantiles.
The source code of the experiments is available online\footnote{\url{https://github.com/QuentinMaz/QD_Based_Testing_RL}}.

\paragraph{Test oracles}

In Lunar Lander, a failure occurs if the lander crashes into the ground or moves outside the viewport.
In Bipedal Walker, failure occurs if the body of the robot collides with the ground.
For the Taxi environment, a fault occurs in case of an illegal action (for instance, dropping the passenger even though the taxi is still empty) or collision (by moving into a wall).

\paragraph{Input/Parameter sampling and mutation}

\begin{figure*}[t]
    \centering
    \includegraphics[width=\textwidth]{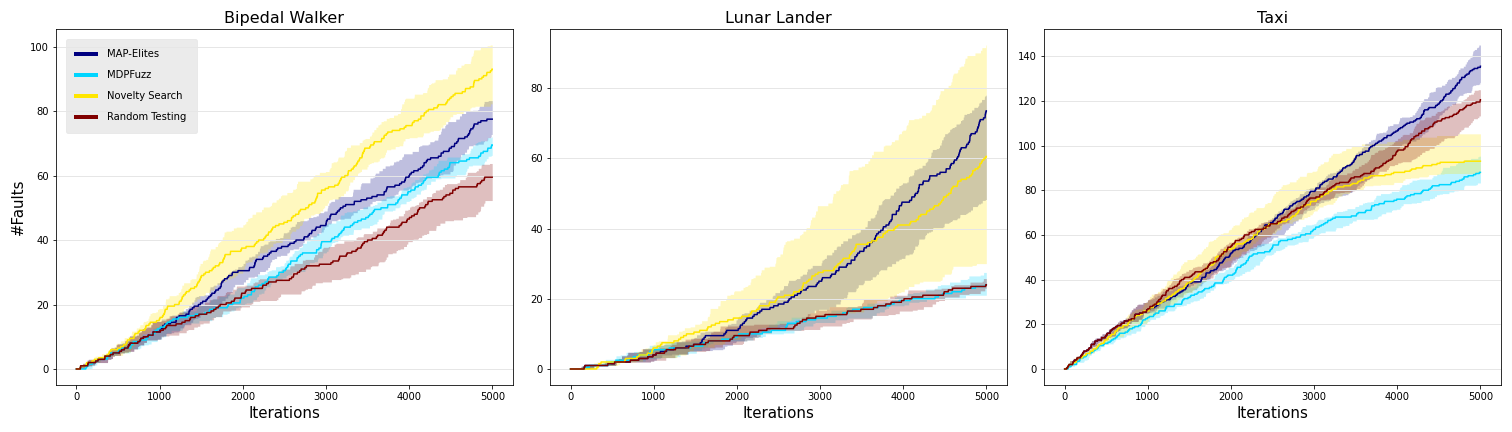}
    \caption{Evolution of the number of fault-triggering solutions found for each framework evaluated. The lines show the median results over 10 executions, and the shaded areas correspond to the first and third quantiles.}
    \label{fig:rq1}
\end{figure*}

For the Lunar Lander environment, the shape of the landscape is fixed by our assumption to consider deterministic environments.
As such, the parameter space is the two-dimensional space $[-1000, 1000]^2$ that describes the possible initial forces applied to the spacecraft.
The mutation operator slightly perturbs the original parameter (clipped, if needed).
For the Bipedal Walker use-case, we follow the experimental settings of MDPFuzz~\cite{10.1145/3533767.3534388} for both the parameter space and the mutation operator.
Precisely, the parameter space encodes the type of obstacles (flat, pit, steps or stump) of the landscape as 15-size vectors whose values $\in [0,1,2,3]$, while the mutation operator randomly changes at least one of those values (clipped if needed).
As mentioned in the aforementioned paper, the obstacles are sufficiently well away from each other so that they can be passed by an optimal policy $\pi^*$ (i.e., all the problems are solvable).
As for the last use-case, the parameter space encodes the initial position of the taxi and the passenger as well as its destination.
The map is static in this environment.
The mutation operator increments or decrements one value (clipped if needed).
For instance, the taxi would now start from a cell in the grid world close to the initial one.

\paragraph{Behaviour Spaces}

All the behaviour spaces studied are two-dimensional spaces.
For the Lunar Lander use-case, we use the behavioural definition proposed by~\cite{pyribs_lunar_lander}.
It describes how the policy lands the spacecraft as its horizontal position and vertical velocity when it first touches the ground.
If the lander moves outside the viewport, we consider the last values observed.
For Bipedal Walker, we follow a previous work that studies policy behaviour for this very same use-case~\cite{10.1145/3377929.3389921}.
In particular, they define a set of hand-designed behaviour descriptors (averaged over the observation state sequence) such as \textit{Distance}, the walker’s position relative to the goal, \textit{Hull angle}, the body's angle of the agent, \textit{Torque}, the force applied to the agent’s hip and knee joints, \textit{Jump}, that describes when both legs are simultaneously not in contact with the ground and \textit{Hip angle/speed}, that are the angle and speed values of the agent’s hip joints, respectively.
We define the behaviour space as the pair of the descriptor \textit{Distance} and \textit{Hull angle}.
The effect of different descriptor pairs as behaviour spaces is addressed in RQ3 (see Subsection~\ref{subsec:rq3}).
As for the last use-case, the behaviour is defined as the two dimensional point that counts the number of actions to 1) pick-up the passenger and 2) to drop them off afterwards.

\paragraph{Models Under Test}

For the Bipedal Walker and Lunar Lander experiments, we use the models freely available on the Stable-Baselines3 repository~\cite{rl-zoo3}.
For the customised Taxi use-case, we train the agent via Q-Learning~\cite{Watkins1992}.

\paragraph{Hyperparameters}

We follow the guidelines indicated in~\cite{10.1145/3533767.3534388} to configure the Gaussian Mixture Models used by MDPFuzz.
As for Novelty Search, it computes the novelty scores as the average distance of the 3 nearest neighbours and the novelty threshold for updating its novelty archive has been set to $t=0.9$ for the Taxi use-case and $t=0.005$ for Bipedal Walker and Lunar Lander.
Similarly to all the previously mentioned experimental parameters, they have been obtained in a prior study.

\subsection{RQ1: Effective Fault Detection}\label{subsec:rq1}

We first investigate the efficiency of the approaches evaluated.
Figure~\ref{fig:rq1} shows the evolution of the number of distinct fault-triggering solutions.

\paragraph{Results}

All the frameworks evaluated find more faults than Random Testing for the Bipedal Walker use-case.
Precisely, at the end of testing, we report an average improvement of 30\% and 56\% for ME and NS, respectively, while MDPFuzz finds 17\% more faults.
Greater improvements of QD optimisation are observed for the Lunar Lander use-case (up to 206\% and 152\%), while MDPFuzz matches the baseline.
We note that in the case of NS, the latter vary a lot (see the shaded areas in Figure~\ref{fig:rq1}).
The results for the last use-case show a different picture though.
Here, only ME beats Random Testing (up to 12\%), while Novelty Search and MDPFuzz find 23\% and 27\% fewer faults, respectively.

\paragraph{Analysis}
Searching for diversity can impede \textit{test efficiency}, as Random Testing outperforms Novelty Search in the Taxi experiments.
Furthermore, NS is the most sensitive framework to randomness, which is likely to be caused by its high dependency to its initial population.
It is therefore difficult to recommend as a general optimiser for efficient QD-based policy testing.
However, accounting for both quality and diversity lets MAP-Elites systematically beat Random Testing, while showing lesser sensitivity to its initialisation.
Besides, MDPFuzz does not seem to be suited for all the use-cases studied, since it is only able to compete with the efficiency of QD-based policy testing on the Bipedal Walker environment.
While being deceptive, those results are not completely surprising.
Indeed, this framework drives its search with Gaussian Mixture Models (GMMs), whose parameters were studied for several applications, including Bipedal Walker.
We therefore suspect that MDPFuzz ends up sharing its results with Random Testing on Lunar Lander because of none-optimal GMM configuration.
As for the Taxi use-case, we recall that the MDPFuzz aims at enabling policy testing solving complex MDPs~\cite{10.1145/3533767.3534388}, which is definitely not the case of this toy problem.

\paragraph{Conclusion} The complexity of dedicated policy testing techniques such as MDPFuzz can hurt their efficiency -- especially for smaller MDPs --, while QD optimisation consistently finds the greatest number of faults in the decision model.
About that, discarding solution quality in favour of novelty can lead to better, yet unstable results.
QD-based policy testing does not come with poorer efficiency (as we expected) but rather with significant increase the number of functional faults found in the model.

\subsection{RQ2: Testing and Fault Diversity}\label{subsec:rq2}

\begin{figure}[t]
    \centering
    \includegraphics[width=\columnwidth]{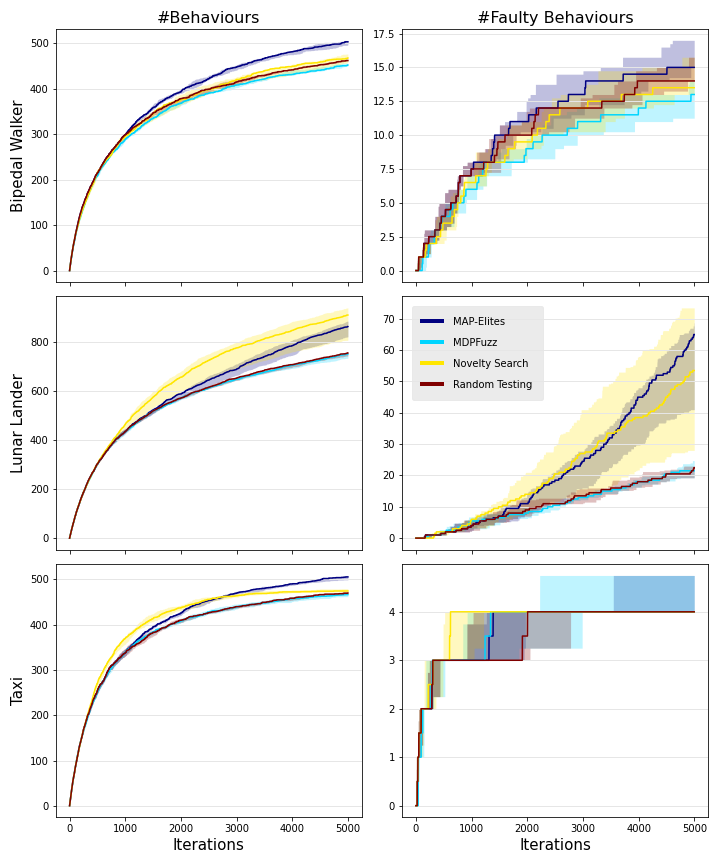}
    \caption{Evolution of the behaviour space coverage over time as the number of behaviour niches (bins) illuminated during testing. In the second column, only bins filled by fault-triggering solutions are counted, i.e., faulty behaviours. The lines show the median results over 10 executions, and the shaded areas correspond to the first and third quantiles.}
    \label{fig:rq2bs}
\end{figure}

\begin{figure}[t]
    \centering
    \includegraphics[width=\columnwidth]{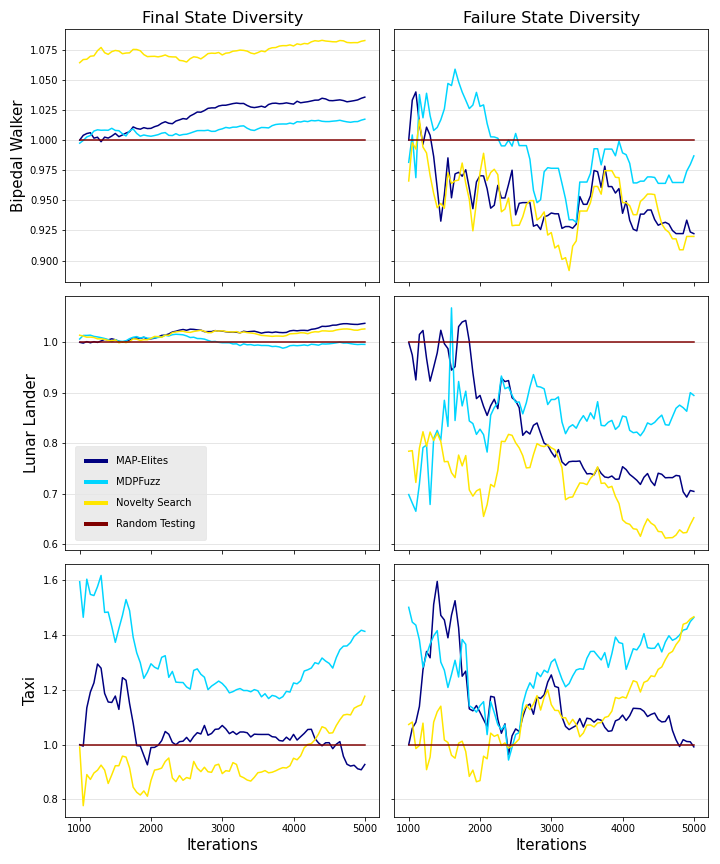}
    \caption{Final state diversity as the average distances of the 3 nearest neighbours. Since their scale depends on the observation space of each use-case, we report the relative performance of the methodologies to Random Testing. The lines show the median results over 10 executions.}
    \label{fig:rq2fs}
\end{figure}

Next, we study how QD optimisation improves diversity in terms of behaviours and final states.
Figure~\ref{fig:rq2bs} shows the behaviour and faulty behaviour coverage, and Figure~\ref{fig:rq2fs} the average distances of the 3 nearest neighbours of the final states.
Since the latter are distances between observation states, Figure~\ref{fig:rq2fs} shows the relative performance of the methodologies evaluated to the Random Testing baseline.

\paragraph{Results}

ME systematically improves behaviour discovery, ranging from 7\% (Taxi) to 14\% (Lunar Lander).
MDPFuzz matches RT's performance for all the use-cases, as Novelty Search does, except for Lunar Lander for which NS covers up to 20\% more behaviour niches (slightly outperforming MAP-Elites).
Regarding faulty behaviours, our results only show significant differences to Random Testing for MAP-Elites and Novelty Search on the Lunar Lander experiments.
Precisely, they stand out after 2000 iterations and end up with 189\% and 138\% more faulty behaviours discovered, respectively.
Similarly to the previous results though, we note that NS's performance vary a lot.
If we now look at final state diversity, we find mixed-bag results.
For Bipedal Walker, we observe that Novelty Search explores around 7.5\% more diverse final states than the random baseline throughout testing.
For Lunar Lander, none of the testing technique significantly improves the baseline's results.
Actually, the failure state distribution tends to be worse, especially for QD optimisation (up to a 30\% decrease).
For the last use-case, MDPFuzz outperforms all the other techniques with significant margins, with 40\% greater distances in the final states (averaged throughout testing) and up to 50\% sparser failure states.
Meanwhile, if QD optimisation does not show significant difference with the baseline for the final state diversity, we observe great, yet unstable improvements in their failure state results (as shown in the second chart of the bottom row in Figure~\ref{fig:rq2fs}).
In particular, Novelty Search catches up with MDPFuzz's performance on the last iterations.

\paragraph{Analysis}

\begin{figure*}[h]
    \centering
    \includegraphics[width=0.9\textwidth]{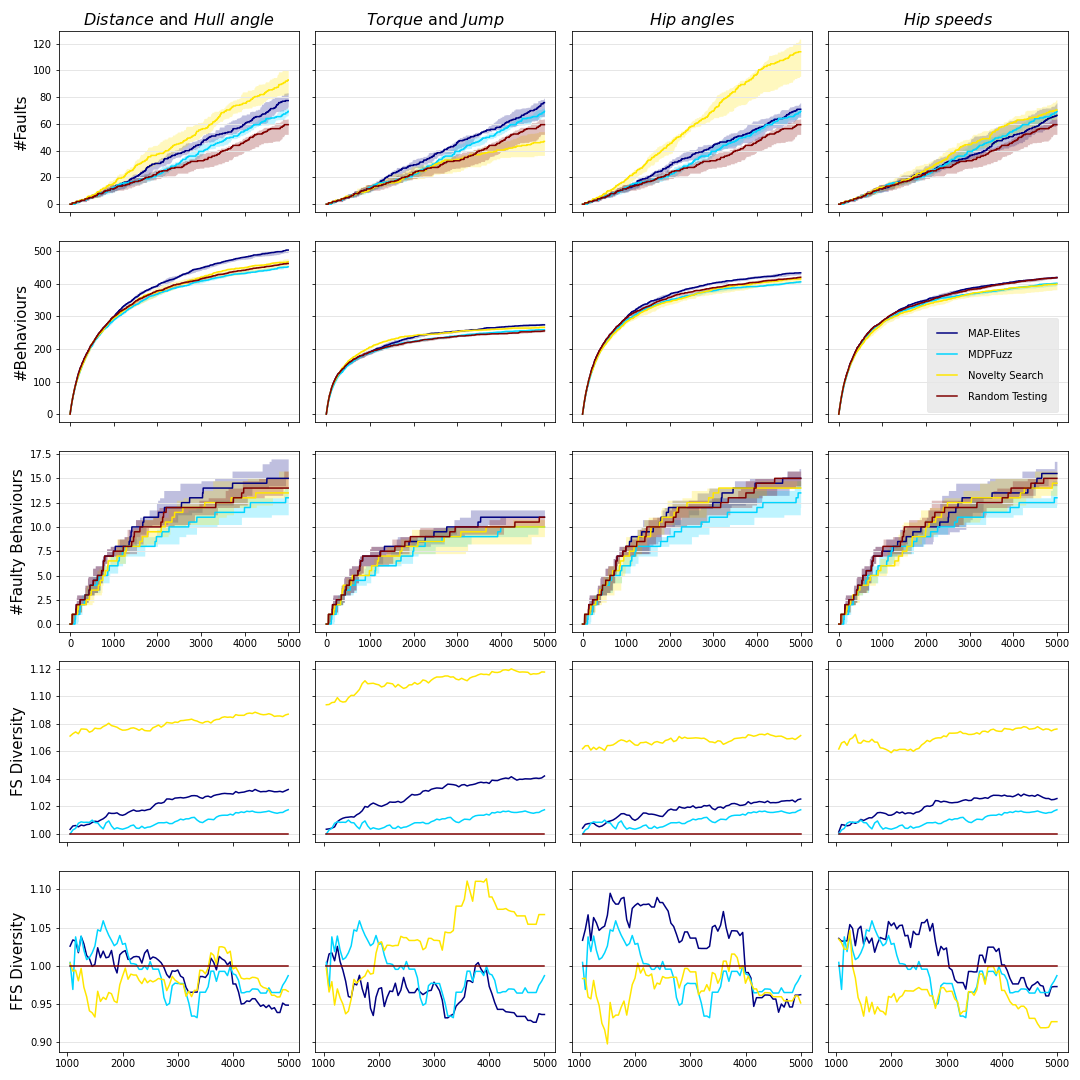}
    \caption{Impact of the behaviour space parameter for the Bipedal Walker experiments. The four spaces are different pairs of hand-designed descriptors studied in~\cite{10.1145/3377929.3389921}. Each column displays the results for a behaviour space. From top to bottom: number of faults, number of behaviours and faulty behaviours, final and failure state diversity (FS and FFS) relative to Random Testing. The results are the medians of 10 executions.}
    \label{fig:rq3}
\end{figure*}

QD-based policy testing discovers more behaviours than both MDPFuzz and Random Testing.
More importantly, in some cases (Lunar Lander), this also applies for fault-triggering solutions.
In other words, QD-based policy testing finds more diverse faults in the model under test than dedicated testing techniques.
It is interesting to note that the SOTA framework MDPFuzz does not cover less behaviours than Random Testing; something that one could have expected.
Complementing the diversity measurement of our behavioural evaluation with final state analysis lets us shade interesting details.
Indeed, discovering more behaviours does not translate to sparser final state distribution, as the two QD-based frameworks either matches or slightly improves Random Testing.
It is especially the case for Lunar Lander, for which both ME and NS significantly improve faulty behaviours coverage (i.e., find diverse faults in the model tested) while showing poor failure state distributions.
As such, the use of behaviour spaces helps policy testing, as simply looking at final states is not enough to cover diverse faults in the model.
If we consider the actual descriptors used to define the behaviour spaces, we can explain the lack of a general correlation between increased behaviour and final state discovery.
Indeed, none of the descriptors solely depends on the termination of an execution to define the behaviour.
For instance, in the Bipedal Walker experiments, the behaviours are averages of some observations' features.

Interestingly though, the descriptors defined for the Lunar Lander use-case let us better understand what type of faults every methodology tends to find.
In particular, behaviours are based on the \textit{first} contact of the spacecraft with the ground; but the policy can later fail if the landing wasn't on the targeted pad (and that the lander has thus to be safely glided towards the latter).
Therefore, by facing the  results of our bi-metric evaluation, an interpretation is that here the QD-based frameworks find solutions for which $\pi$ lands the spacecraft at various positions (i.e., diverse behaviours) but later fails the task by colliding the lander's body on the same edge of the landscape (i.e., dense failure states), while MDPFuzz either finds slightly more different edges (around 10\% more distributed failure states) but the policy actually lands at similar positions (i.e., poor faulty behaviour discovery).

\paragraph{Conclusion}

The ability of QD optimisation to find diverse high-quality solutions applies for policy testing, that is, fault-triggering inputs that exercise the model such that it fails with varied behaviours.
As found in the previous analysis, discarding solution quality leads to unstable results, which is fixed by more balanced Quality Diversity optimisers such as MAP-Elites.

\subsection{RQ3: Behaviour Space Impact}\label{subsec:rq3}

In this last research question, we investigate the effect of different behaviour spaces on the proposed QD-based policy testing framework.
As previously mentioned, \citeauthor{10.1145/3377929.3389921}~\cite{10.1145/3377929.3389921} define several descriptors to characterise the behaviour of a policy.
We use the ones introduced in Subsection~\ref{subsec:impl} to define three additional behaviour spaces (as descriptor pairs) to study how the results of our appraoch can differ.
Figure~\ref{fig:rq3} summarises our findings, where each column corresponds to a behaviour space.
Note that the first column corresponds to the results found above.
In the following, we analyse the impact on each metric.

\paragraph{Test efficiency}

ME improves the number of faults found compared to Random Testing, regardless of the behaviour space used, to a extent that can decrease to 11.5\% (down from 30\%, the first column on Figure~\ref{fig:rq3}).
However, Novelty Search shows significantly higher sensitivity to the behaviour space.
In particular, NS can almost double the number of faults found by Random Testing with the third behaviour definition (90\%), but has around 20\% poorer performance with an inappropriate space (as shown in the second column of Figure~\ref{fig:rq3}).
The drastic difference in behaviour sensitivity of the two implementations again lies in the balance between how quality and diversity of solutions are accounted.
Since MAP-Elites mostly relies on quality (by focusing on mutating its \textit{elites}), the results are steadier and less sensitive to their exact behaviour.
In other words, as long as the behaviour space is able to capture diversity -- which is the case here, as all the spaces are based on meaningful, hand-designed descriptors -- MAP-Elites seems to be an efficient optimiser for QD-based policy testing.
On the other hand, by discarding quality to only account for behaviour novelty, Novelty Search becomes more sensitive to the behaviour space used.

\paragraph{Behaviour Diversity}

The number of behaviours discovered by all the methodologies are for the most part steady across the spaces evaluated, which match the performance of the baseline.
We only observe ME discovers around 8\% more behaviours with two of the spaces (first and third columns in Figure~\ref{fig:rq3}).
As for the diversity of faulty behaviours, we report no significant change compared to our previous findings, that is: all frameworks share similar figures.
One can possibly note MDPFuzz has consistent lower numbers (10\% in averaged over the spaces).
While QD-based policy testing never impedes behaviour discovery, its ability to significantly improve the random baseline depends on the behaviour space.
For instance, we find that the behaviour of the policy tested is best captured with the first pair of descriptors.

\paragraph{Final State Diversity}

The behaviour space definition does not affect the sparseness of the final states, which is little surprising since behaviours are computed as feature averages.
Similarly, the relative performance of the methodologies to the baseline does not significantly fluctuate.
In particular, Novelty Search shows a 8\% to 12\% smoother state distribution.
Finally, if we only consider failure states (bottom row), we can see that there is no general trend in the results neither, with minor fluctuations around the baseline lower than 11\% throughout testing.

\subsection{Summary}

Our first and most important conclusion is that QD manages to find diverse faults in the decision model, despite the simplicity of our approach.
In particular, it shows that complex, dedicated policy techniques such MDPFuzz are not always needed.
Second, we systematically observe a lack of consistency in the results of our framework optimised with Novelty Search, especially when running close to our set test budget.
As such, despite its outstanding performances for some cases, we recommend using our proposal with QD optimisers that account for quality and diversity in more balanced ways.
Finally, the selection of the precise behaviour space can boost the performance of QD-based policy.
We observe this for the Bipedal Walker use-case, where a well-chosen behaviour space substantially increases the number of faults detected.
At the same time, the other behaviour spaces are all competitive and do not negatively impact the ability to reveal diverse faults.

\section{Threats to Validity}\label{sec:threats}
In the following, we discuss the limitations of our proposal as well as the threats of our experimental evaluation.

\paragraph{External Threats}
A first threat to our evaluation is the policy under test used, that we mitigate by using the same model as previous works~\cite{10.1145/3533767.3534388}.
Similarly, there is some inherent bias to the results from the selected use-cases.
To that regard, we consider three environments of various natures, two of them having already been studied in the QD and RL literature.
Finally, we only evaluate one SOTA policy testing technique, namely MDPFuzz.
Given the space limitation, we decided to prioritise the number of use-cases, since this work primarily introduces Illumination optimisation in policy testing.

\paragraph{Construction Threats}
In this first work, we assume to have deterministic executions with fixed randomness effects.
While deterministic executions are a common approach for testing policies and their deployment, the challenge of handling MDP stochasticity lies in the fact that multiple executions of a particular solution would generate different trajectories and thus, possibly different behaviours.
Similarly, some of these executions might fail depending on the selected policy actions or state transitions in the MDP.
We acknowledge that it is thus an important challenge in QD-based policy testing, that needs to be addressed in future work.
Practically speaking, by fixing the random effects we reduce the search space to a subset of inputs that reveal a fault with the given random seed.
We thereby limit the experimental evaluation to a subset of all possibly detectable faults in the policy.

\section{Conclusion}
\label{sec:conclusion}

This work introduces Quality Diversity (QD) for policy testing.
QD is a flexible, black-box optimisation framework that optimises a population of individuals by considering both their behaviour, i.e. how they solve a given problem, and their quality, i.e. how well they solve it.

We illustrate how to adapt QD to policy testing and propose the first formulation of diversity-oriented policy testing as an Illumination task.
Precisely, we characterise test inputs with the behaviour of the policy under test, that is, how it solves/fails the test case.
We implement our QD-based testing framework with two commonly used QD optimisers of different paradigms: the elitist MAP-Elites and the divergent Novelty Search algorithms.
We perform experiments on three use-cases from the reinforcement learning literature, and compare QD-based testing to state-of-the-art policy testing and random testing as the state-of-the-practice.
Our results show that QD optimisation, while being a conceptually straightforward and easy-to-apply approach, not only improves fault diversity but also fault detection.
We further assess the impact of the behaviour space definition, what we consider as the most decisive parameter of our approach.
With this first work, we open a new application area for Quality Diversity.

In future work we will address the inclusion of generic respectively learned behaviour spaces to reduce the initial effort to setup QD and the handling of stochastic MDPs in the search space to further guide the search.

\section*{Acknowledgements}
This work is funded by the Norwegian Ministry of Education and Research, the Research Council of Norway under grant number 324674 (AutoCSP), and is part of the RESIST\_EA Inria-Simula associate team.
\bibliographystyle{ACM-Reference-Format}
\bibliography{refs}
\end{document}